# Multiple polar and non-polar nematic phases

S. Brown[a], E. Cruickshank[a], J.M.D. Storey[a], C.T. Imrie[a], D. Pociecha[b], M. Majewska[b], A. Makal[b] and E. Gorecka [b] *

[a]   Department of Chemistry, University of Aberdeen, Old Aberdeen AB24 3UE, U.K.

[b]   Department of Chemistry, University of Warsaw, Zwirki i Wigury 101, 02-089 Warsaw, Poland

**Abstract:** Liquid crystal materials exhibiting up to three nematic phases are reported. Dielectric response measurements show that while the lower temperature nematic phase has ferroelectric order and the highest temperature nematic phase is apolar, the intermediate phase has local antiferroelectric order. The modification of the molecular structure by increasing the number of lateral fluorine substituents leads to one of the materials showing a direct isotropic-ferronematic phase transition.

## Introduction

The nematic phase is the least ordered liquid crystalline phase in which the rod-like molecules are statistically oriented along a common direction, called the *director*, **n**, whereas their centers of mass are distributed randomly, and thus, the phase has a fluid character. The *director* is a unit vector having inversion symmetry, i.e. **n = −n**, so the phase is non-polar. This picture of the nematic phase has been accepted for many years, but it is worth noting that nearly a hundred years ago, and at the very beginning of liquid crystal research, an alternative model of the nematic phase was also considered; Born discussed the ferroelectric nematic phase in which the molecular dipole moments are also ordered [1]. He argued that to obtain a polar fluid, the interactions between the dipole moments should be strong enough to withstand thermal fluctuations. Although Born's model was subsequently rejected on experimental grounds, there is no fundamental reason why a ferroelectric nematic phase should not exist, and such polar liquids were predicted also by theoretical modelling [2]. Ferroelectric ordering in liquids is absolutely fundamental not only in chemistry and physics, but could also have far-reaching implications in the biological sciences. For example, it was predicted that the ferroelectric nematic phase, in order to reduce electrostatic energy, will twist giving a polar cholesteric phase, and this spontaneous chirality would be controlled through steric and electrostatic interactions between achiral molecules [3]. Given that chirality is widely believed to hold the key to an understanding of the origins of life, and liquid crystals are ideal systems to study chirality, its origins and chirality propagation, polar ordering in liquid crystal phases could have far reaching significance.

In 2017, two independent reports were published claiming the discovery of a new type of nematic phase for two quite different molecular structures, RM734 and DIO (Figure 1), [4] which was later claimed to have ferroelectric properties, and termed the $N_F$ phase [5]. These claims have again raised the question as to whether polar forces are sufficient to give a stable fluid ferroelectric phase. Prior to this finding, the polar order observed in fluids resulted from interactions other than dipolar forces. The first ferroelectric liquid crystalline phase was discovered almost fifty years ago by R.B. Mayer et al. using symmetry arguments to predict the polar properties of tilted smectic phases, and specifically that dipole order becomes possible if the symmetry of the structure is sufficiently reduced by constructing the tilted smectic phase from chiral molecules. [6] Some twenty-five years later, H. Takezoe and colleagues demonstrated an alternative strategy for obtaining polar properties in smectics, in which the ordering of the dipoles was a consequence of the restricted rotation of molecules arising from their bent shape [7]. In both types of polar smectic phases, the dipole order is a consequence of steric interactions, and not dipolar interactions, thus they are considered as improper ferroelectrics. The first proper ferroelectric liquid crystal phase was reported for the polymer called Vectra (polypeptide PBMLG) which forms a lyotropic cholesteric phase in benzyl alcohol; the ferroelectric properties of this phase were established using switching current and second harmonic generation (SHG) methods [8].

Although the recent discoveries of polar nematic phases involved chemically rather different compounds[4,5], their molecular architectures and those of other examples that exhibit this new polar nematic phase do share some common features. Notably, these systems have a large longitudinal dipole moment, and often have a nitro terminal group. It has been suggested that the nitro group reduces the tendency for these molecules to form antiparallel dimers. Secondly, the molecules tend to be wedge-shaped. These structural requirements for the observation of the $N_F$ phase are consistent with computer simulations of tapered Gay-Berne particles with a longitudinal dipole moment that revealed $N_F$ behaviour [9]. However, since only a small handful of compounds that show this new polar nematic phase have been reported to date, it remains far from clear whether the polar order in the $N_F$ phase is a direct consequence of dipole-dipole interactions, or if it is induced/enhanced by steric interactions and is associated with 1D or 2D splay deformations [10].

Here we report the further characterisation of the two prototypical ferroelectric nematogens, RM734 and DIO, and compare their properties to two new materials, see Table 1. The syntheses of the materials and their intermediates is described fully in the supporting information.

## Results and Discussion

**Studied materials:** A better understanding of the polar nematic phase requires an increase in the library of compounds that show such a phase. Here we investigate the properties of four materials: compound **1** (RM734 [5b]) is a "prototype" ferroelectric nematogen; compounds **2** and **3** are new derivatives of compound **1**, laterally substituted with one and two F atoms, respectively; compound **4** (DIO [4b]) has different mesogenic core and contains the 1,3-dioxane unit (Fig. 1). The molecules of all the studied compounds have a strong longitudinal dipole moment, and a conical shape due to the lateral substituents. The calculated dipole moments (see SI) were as follows: 11.3950 D for compound **1** (RM734), 12.0496 D for compound **2** and 13.1345 D for compound **3** and 8.9214 D for compound **4** (DIO).

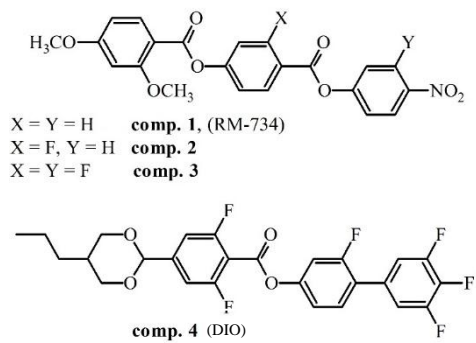

**Figure 1.** *Molecular structures of the studied materials.*

**Table 1.** Phase transition temperatures (in °C) and enthalpy changes (in parentheses, in kJ mol$^{-1}$) determined by DSC for the studied compounds; m.p. stands for melting point, phase sequence is given on cooling scans.

| Comp. | m.p. | Phase sequence |
|---|---|---|
| 1 | 139.3 (34.1) | Iso 187.0 (0.6) N 130.7 (0.6) $N_F$ |
| 2 | 160.7 (54.1) | Iso 164.7 (0.7) N 142.5 (1.4) $N_F$ |
| 3 | 151.1 (53.1) | Iso 139.1 (4.4) $N_F$ |
| 4 | 95.9 (28.9) | Iso 174.0 (0.6) N 83.9 (0.005) $N_x$ 68.1 (0.2) $N_F$ |

**Phase sequence:** Comparing the properties of materials **1-3**, it is clear that increasing the number of lateral fluorine substituents reduces the clearing point (Table 1), which is expected for molecules with decreasing shape anisotropy, but surprisingly, strongly increases the stability of the $N_F$ phase relative to the N phase. As a result, for compound **3** a direct transition from the isotropic liquid to the polar $N_F$ phase is observed. Such a direct transition has so far been reported only for one compound[5e]. For material **4** an additional nematic phase, $N_x$, is observed between the N and $N_F$ phases[4b], in a temperature range of 17 K.

For all the studied materials, in cells with planar anchoring (HG cells) with anti-parallel rubbing, in the N phase a uniform texture was observed (Fig. 2a), which is retained just after the transition to the $N_F$ phase (Fig. 2c). On further cooling, twisted states are usually formed a few to several degrees below the phase transition (Fig. 2d-f), depending on the cooling rate and cell thickness. The twist is induced to connect molecules oriented with opposite polarization vectors on the lower and upper cell surfaces; the anti-parallel polarization orientation on the surfaces is due to the anti-parallel rubbing of the polymer used as a surfactant (inset in Fig. 2). In the cells with homeotropic anchoring condition (HT cells), a birefringent, schlieren texture is observed in the nematic phase (Fig. S1 and S2), apparently the polymer anchors the director with some tilt from the surface normal. Both, two- and four-brush point defects are present. In the $N_F$ phase, additional defects arise (Fig S1 and S2) that divide regions with different orientation of polarization in the cell bulk; the appearance of domains with different polarization is driven by director splay deformation which is necessary to connect the polar vectors oriented inside (or outside) on both, the lower and upper cell surfaces. For compound **4** exhibiting the additional $N_X$ phase, the optical textures of the N and $N_X$ phases are similar, the phase transition is observed as a weak wave-like front going across the cell, the cessation of director fluctuations (flickering) and the appearance of zig-zag defects (Fig. 2b).

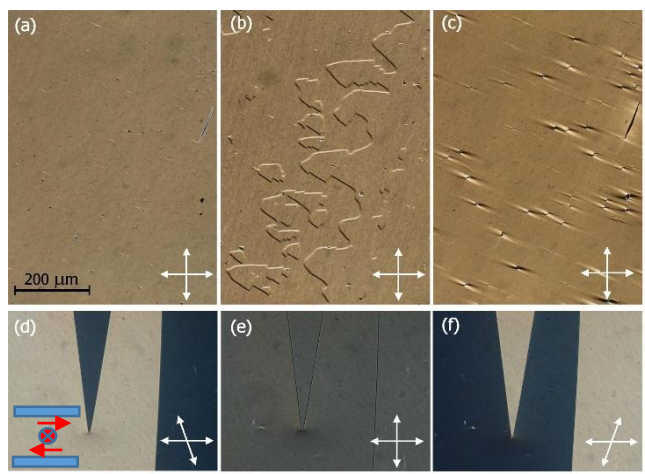

**Figure 2.** *Optical textures of compound **4** in (a) N, (b) $N_x$ and (c) $N_F$ phases in 5-micron-thick planar cell. In the $N_F$ phase, twisted states are formed a few degrees below the phase transition (d-f). In the inset, a schematic drawing of one of the twisted states in which the polarization is anchored at the cell's surfaces by antiparallel rubbed polymer layer, the red arrows are polarization vectors.*

The narrower the range of the nematic phase, the more first-order in nature the transition to the $N_F$ phase becomes, which is reflected in the calorimetric measurements; for example, the thermal effect accompanying the N-$N_F$ phase transition for material **2** with a narrower N phase range, is twice as strong than for compound **1** (Tab. 1). For material **3**, the Iso-$N_F$ transition is accompanied by a relatively strong enthalpy change of ~10 J/g, much higher than that normally observed at the Iso-N phase transition. Seemingly, the N-$N_F$ or Iso-$N_F$ phase transition, which includes not only the changes of the orientational order of the molecular axes but also the ordering of the dipole moments, requires an additional contribution of entropy. For material **4**, calorimetric measurements in heating and cooling runs, consistently show the existence of three nematic phases below the isotropic phase, the N-$N_x$ phase transition is accompanied by a very small thermal effect (~0.01 J/g), much smaller than thermal effect at the $N_x$-$N_F$ (~1 J/g) phase transition (Fig. S3). This suggests that $N_X$ is more similar to the N phase than to the $N_F$ phase and probably does not show long range dipole ordering. This view is also consistent with the observed texture changes: in the $N_x$ phase in an HT cell, the point defects s=1/2 are preserved (Fig. S2), that are not allowed in the polar phase. Birefringence measurements are consistent with the calorimetric data:

specifically, the transition from the N to $N_F$ phase is first order, and is accompanied by a jump in $\Delta n$ (Fig. 3), which for both materials **1** and **2** corresponds to an increase in the orientational order parameter $\Delta S \sim 0.1$. Apparently, the formation of long-range polar order increases the orientational order of the molecules. Several degrees below the transition to the $N_F$ phase, the ordering of the long molecular axes is nearly perfect, and parameter S reaches 0.9 (Fig. 3). For compound **4**, in the $N_X$ phase the birefringence shows only a slight change in comparison with the N phase (Fig. 3), and in the $N_X$ phase grows slightly faster than in the N phase. There is a clear step-like increase in birefringence, thus also in the order parameter S, at the $N_X$-$N_F$ phase transition.

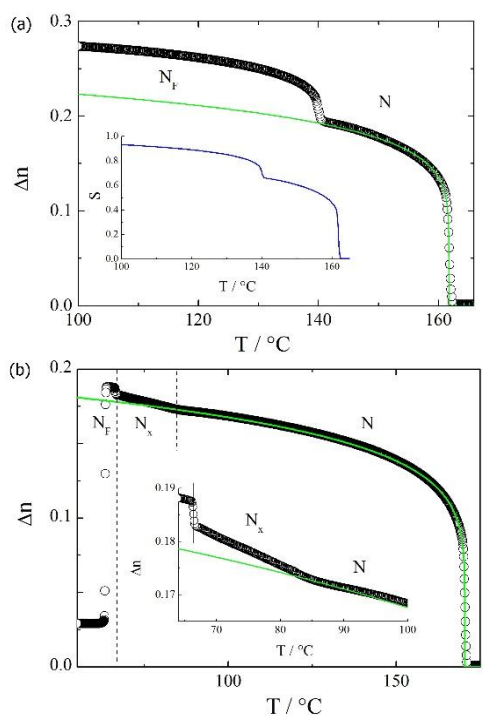

**Figure 3.** *Birefringence vs. temperature for (a) compound **2** and (b) compound **4**. The strong decrease of the measured $\Delta n$ in the $N_F$ phase a few K below the $N_X$-$N_F$ phase transition in (b) is due to the formation of twisted states in the cell. In the inset of (a) the order parameter vs. temperature; in the inset of (b), an expanded section of the main graph showing the temperature range close to the N-$N_X$ phase transition. Green lines show the fitting to critical dependence with $\Delta n_{max}$=0.294 for compound **2** and $\Delta n_{max}$=0.223 for compound **4**.*

The lack of positional order of the molecules in the temperature ranges of the liquid crystalline phases was confirmed by x-ray diffraction studies. In all the nematic phases we observed diffuse diffraction signals corresponding to the full length of the molecule (Fig. S4), and no evidence for the antiparallel pairing of molecules, typical for mesogens with strong polar end groups, was found.

**Polar properties:** For all the materials the dielectric response strongly depends on the cell thickness and surface anchoring conditions (HG or HT) used. However, the measured permittivity value should be treated with some caution - the thin (~10 nm) polyimide aligning layers present at the cell surfaces act as additional high capacitance being in series circuit with the capacitor filled with LC sample. Thus, for studied here materials with very high permittivity the measured equivalent capacity of the circuit might be considerably lower than the actual capacity of LC layer.

The dielectric measurements performed in planar cells show the dielectric relaxation mode only in the $N_F$ phase, and for compound **3** a strong dielectric mode appears directly below the transition from the isotropic liquid (Fig. S5). In HT cells (anchoring director at some pretilt angle to the surface normal) or in cells without additional surface treatment (therefore random orientation of director), in the $N_F$ phase the dielectric mode is much stronger; for example, for compound **2**, the dielectric strength of the mode reaches almost 10000 (Fig. S6).

Furthermore, in HT cells there is also a clear dielectric relaxation mode in the N phase. Although the N-$N_F$ phase transition is first order in nature, some critical decrease in the relaxation frequency of the mode and an increase in its strength were observed in the N phase as the system approaches the $N_F$ phase, indicating an increase in the correlation of dipole motions (Fig. S6).

The absence of a dielectrically active relaxation mode in the N phase in the cells with planar anchoring (therefore with dipoles parallel to the surface) and the presence of a clear mode in HT cells indicates that the dipole fluctuations are related to the collective rotation of molecules about their short molecular axis.

The most interesting properties are observed for material **4** (Fig. 4). The dielectric measurements in the HT cells showed a single relaxation mode in the N phase, the frequency of which decreases and its strength increases on cooling, indicating growing polar order; two weak modes in the $N_x$ phase, one which is continuous from the N phase, and the other with a much higher relaxation frequency; and a single, strong relaxation process in the $N_F$ phase with a relaxation frequency ~2 kHz, similar to that seen in the other studied compounds (Fig. 4). When a bias electric field is applied across the cell in the N phase near to the transition to the $N_X$ phase, the mode strength increases significantly, and its relaxation frequency slightly reduces. Apparently, the electric field changes the orientation of the molecules, a large fraction of the molecules (which in the ground state splay across the HT cell) become oriented with their long axes perpendicular to the surface, and as a result the fluctuations of their dipoles become more visible in dielectric spectroscopy. In the $N_X$ phase, above some critical threshold (~0.12 V $\mu m^{-1}$), the bias field suppresses the higher frequency mode and simultaneously increases the strength of the lower frequency mode, only slightly lowering its frequency (Fig. 4c,d,f). From dielectric spectroscopy measurements performed under a sufficiently strong bias field (0.25 V $\mu m^{-1}$), it is quite obvious that the electric field transforms the $N_X$ phase into a phase with strong polar fluctuations that develops almost continuously across the N-$N_X$ phase transition (Fig. S7). A possible explanation for this is that while in the N phase, the ferroelectric fluctuations dominate, in the $N_X$ phase both local antiferroelectric and ferroelectric fluctuations coexist, and under the electric field, antiferroelectric order is destroyed and the phase is transformed into the nematic phase with short range ferroelectric order. Alternatively, the $N_X$ phase might have some polar order but there is an additional periodic director structure of nm scale that suppresses dielectrically active fluctuations but induces some additional mode due to the elastic deformation of the structure. However, measurements of the spontaneous electric polarization performed for material **4** (Fig. 5) show non-zero $P_s$ values only in the $N_F$ phase, which critically decreases on heating towards the $N_X$ phase. The non-zero electric polarization detected in the $N_X$ phase is a result of the high non-linear susceptibility to an electric field in this phase. The size of polar domains with uniformly aligned polarization was determined by piezoresponse force microscopy (PFM) technique to be more than 10 μm (Fig. S9).

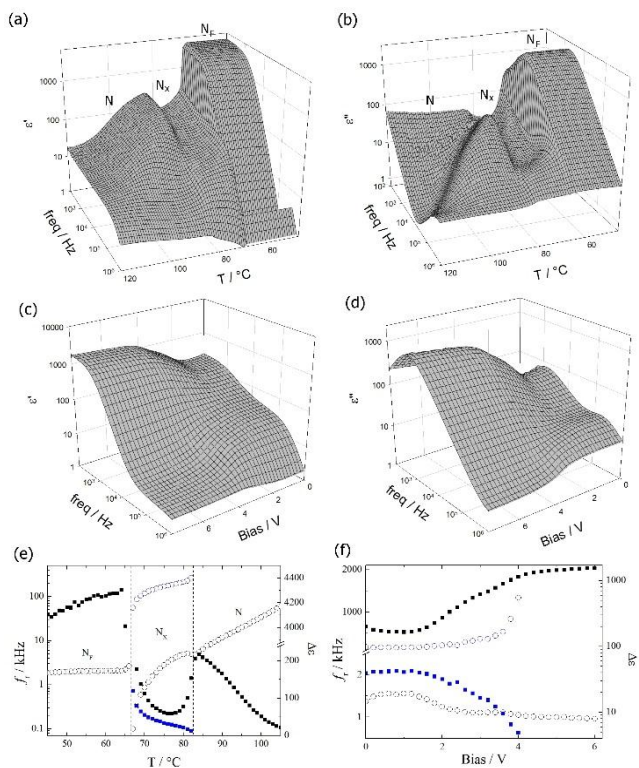

**Figure 4.** Dielectric dispersion *for compound 4 measured in 20-μm-thick cell with homeotropic anchoring: (a) real and (b) imaginary part of dielectric susceptibility vs. temperature and frequency; (c) real and (d) imaginary part of dielectric susceptibility vs. bias voltage and frequency in the $N_X$ phase (T = 70 °C). Relaxation frequency (open circles) and dielectric mode strength (solid squares) evaluated from above data by fitting to Cole-Cole formula: (e) vs. temperature and (f) vs. bias voltage in $N_x$ phase.*

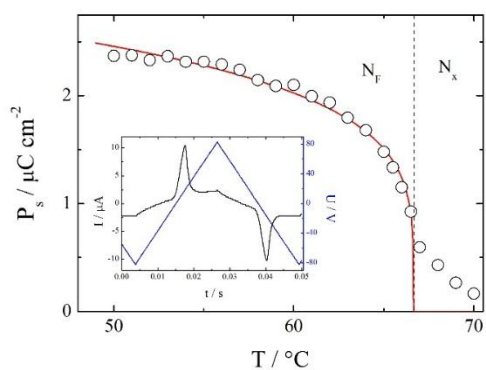

**Figure 5.** *Temperature dependence of electric polarization, $P_s$, for compound 4. In the inset, switching current profile recorded at T=60 °C.*

It should be noted that the electric switching current in the $N_F$ phase is accompanied by clear optical switching, director orientation switches between splay state at zero applied voltage and uniform states along applied field. In the $N_X$ phase only some non-characteristic optical changes were detected under the application of ac voltage.

The splay elastic constant was measured for compound **4**, the $K_{11}$ slightly decreases on cooling in the nematic phase (from ~4 to ~1.5 pN), and at the N-$N_x$ phase transition there is a stepwise increase of $K_{11}$ to ~10 pN. In the $N_F$ phase the splay elastic constant cannot be determined as no influence of electric field on dielectric constant was found, up to the high voltage that destroys sample alignment.

Finally, it should be pointed that compound **4** crystallizes in the non-polar space group of $P2_1/c$ symmetry (see SI).

## Conclusion

Decreasing the shape anisotropy by substituting the mesogenic core of the prototype RM734 molecule with fluorine atoms stabilizes the polar $N_F$ phase (generally a larger dipole moment gives a higher transition temperature to the $N_F$ phase) and destabilizes the N phase. For one of the studied materials, a direct Iso-$N_F$ transition was found and only the second example reported to date. The re-investigation of compound **4** (DIO), showed three nematic phases with a small but distinct transition enthalpy between the N and $N_X$ phases. The intermediate phase that appears between the $N_F$ and N, is apolar, and appears to have the same macroscopic symmetry as the N phase, but differs in short range fluctuations. Such a nematic-nematic phase transition driven by fluctuations are not very common but have been reported, for example, for re-entrant nematics [11]. Under an external electric field, the aniferro-type fluctuations are supressed while ferro-type ones are enhanced. The ferro-fluctuations critically grow with decreasing temperature. The strength of the mode in the $N_F$ phase strongly depends on the cell geometry in which the dielectric measurements are performed.

## Acknowledgements

M. M., D. P. and E. G. acknowledge the support of the National Science Centre (Poland) under the grant no. 016/22/A/ST5/00319. C.T.I. and J.M.D.S. acknowledge the financial support of the Engineering and Physical Sciences Research Council [EP/V048775/1].

**Keywords:** ferroelectric nematic• dielectric constant

# Supporting Information

# Table of Contents

## Experimental Procedures

### Synthesis

**Reagents**

All reagents and solvents that were available commercially were purchased from Sigma Aldrich, Fisher Scientific or Fluorochem and were used without further purification unless otherwise stated.

**Thin Layer Chromatography**

Reactions were monitored using thin layer chromatography, and the appropriate solvent system, using aluminium-backed plates with a coating of Merck Kieselgel 60 F254 silica which were purchased from Merck KGaA. The spots on the plate were visualised by UV light (254 nm) or by oxidation using either a potassium permanganate stain or iodine dip.

**Column Chromatography**

For normal phase column chromatography, the separations were carried out using silica gel grade 60 Å, 40-63 μm particle size, purchased from Fluorochem and using an appropriate solvent system.

**Structure Characterisation**

All final products and intermediates that were synthesised were characterised using $^1$H NMR, $^{13}$C NMR and infrared spectroscopies. The NMR spectra were recorded on either a 400 MHz Bruker Avance III HD NMR spectrometer, or a 300 MHz Bruker Ultrashield NMR spectrometer. The infrared spectra were recorded on a Thermal Scientific Nicolet IR100 FTIR spectrometer with an ATR diamond cell.

**Purity Analysis**

In order to determine the purity of the final products, high-resolution mass spectrometry was carried out using a Waters XEVO G2 Q-Tof mass spectrometer by Dr. Morag Douglas at the University of Aberdeen.

**4-[(Benzyloxy)carbonyl]phenyl 2,4-dimethoxybenzoate**

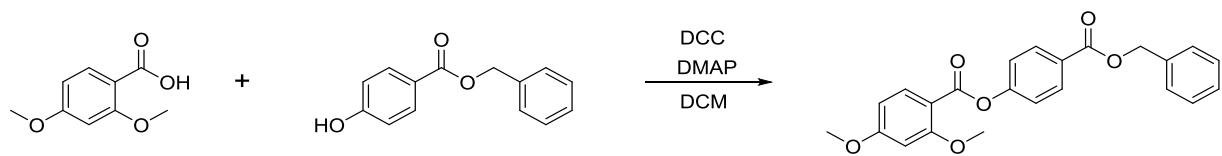

*Scheme 1 - Synthesis of 4-[(benzyloxy)carbonyl]phenyl 2,4-dimethoxybenzoate*

To a pre-dried flask flushed with argon and kept in an ice bath in order to maintain the temperature at 0 °C, 2,4-dimethoxybenzoic acid (4.00 g, 0.0220 mol), benzyl 4-hydroxybenzoate (5.52 g, 0.0242 mol) and 4-dimethylaminopyridine (0.342 g, 2.80×10$^{-3}$ mol) were added. The solids were solubilised with dichloromethane (150 mL) and stirred for 10 min before *N,N'*-dicyclohexylcarbodiimide (5.78 g, 0.0280 mol) was added to the flask. The temperature of the reaction mixture was increased to room temperature and the reaction was allowed to proceed overnight. The extent of the reaction was monitored by TLC using 20 % 40:60 petroleum ether and 80 % ethyl acetate as the solvent system (RF value quoted in the product data). The white precipitate which formed was removed by vacuum filtration and the filtrate collected. The collected solvent was evaporated under vacuum to leave a white solid which was recrystallised from hot ethanol (100 mL).

Yield: 5.82 g, 67.4 %. RF: 0.32. MP: 92 °C

IR ($v_{max}$/cm$^{-1}$): 2940 (C-H), 1749 (C=O ester)

$^1$H NMR (400 MHz, CDCl$_3$): 8.13 (2 H, d, J 8.6 Hz, Ar-H), 8.07 (1 H, d, J 8.7 Hz, Ar-H), 7.41 (5 H, m, Ar-H), 7.29 (2 H, d, J 8.6 Hz, Ar-H), 6.56 (1 H, dd, J 8.7 Hz, 2.3 Hz, Ar-H), 6.53 (1 H, d, J 2.3 Hz, Ar-H), 5.37 (2 H, s, (C=O)-O-C<u>H</u>$_2$-), 3.92 (3 H, s, Ar-O-C<u>H</u>$_3$), 3.89 (3 H, s, Ar-O-C<u>H</u>$_3$)

$^{13}$C NMR (101 MHz, CDCl$_3$): 165.83, 165.22, 162.97, 162.41, 155.00, 136.02, 134.59, 131.19, 128.62, 128.27, 128.18, 127.28, 122.03, 110.68, 104.90, 99.02, 66.76, 56.04, 55.62

Data consistent with reported values.[1,2]

### 4-(2,4-Dimethoxybenzoyloxy)benzoic acid

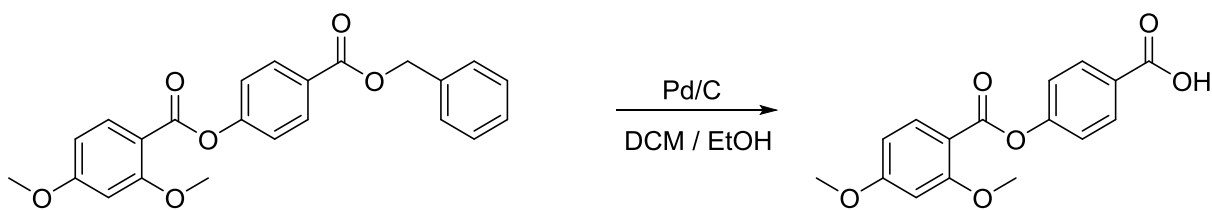

*Scheme 2 - Synthesis of 4-(2,4-dimethoxybenzoyloxy)benzoic acid*

To a pre-dried flask flushed with argon, 4-[(benzyloxy)carbonyl]phenyl 2,4-dimethoxybenzoate (2.50 g, 6.37×10-3 mol) was dissolved in a mixture of dichloromethane (50 mL) and ethanol (50 mL), and added with stirring. The mixture was sparged with argon and 5% Pd/C catalyst (0.490 g, 4.60×10-3 mol) was added. The argon atmosphere was evacuated under vacuum and replaced by hydrogen gas. The reaction was allowed to proceed for 4 h at room temperature, with the extent of the reaction monitored by TLC using ethyl acetate as the solvent system (RF value quoted in the product data). The hydrogen gas, after the reaction was completed, was evacuated under vacuum and the flask was purged using argon. The mixture was filtered through Celite, and the collected solvent was evaporated under vacuum to leave a white solid which was carried forwards without any further purification.

Yield: 1.68 g, 87.1 %. RF: 0.029

$T_{Crl}$ 203 °C $T_{NI}$ (163 °C)

IR ($v_{max}$/cm$^{-1}$): 3382 (OH), 1738 (C=O ester)

$^1$H NMR (400 MHz, $C_2D_6OS$): 13.01 (1 H, br, (C=O)-OH), 8.01 (2 H, d, J 8.3 Hz, Ar-H), 7.96 (1 H, d, J 8.7 Hz, Ar-H), 7.33 (2 H, d, J 8.3 Hz, Ar-H), 6.71 (1 H, d, J 2.4 Hz, Ar-H), 6.67 (1 H, dd, J 8.7 Hz, 2.3 Hz, Ar-H), 3.87 (6 H, s, Ar-O-CH$_3$)

$^{13}$C NMR (101 MHz, $C_2D_6OS$): 167.14, 165.38, 163.02, 162.18, 154.75, 134.45, 131.30, 128.57, 122.75, 110.38, 106.18, 99.49, 56.48, 56.19

Data consistent with reported values.[1,2]

### Compound 1: 4-[(4-Nitrophenoxy)carbonyl]phenyl 2,4-dimethoxybenzoate (RM734)

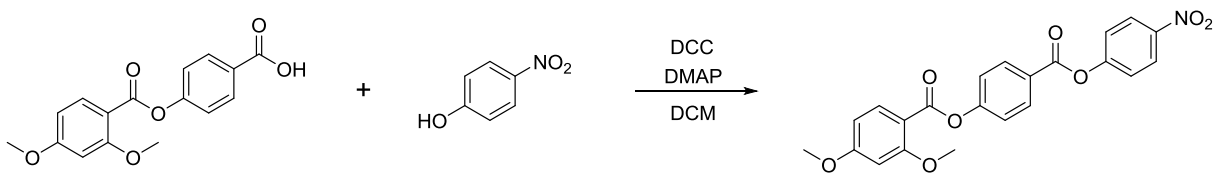

*Scheme 3 - Synthesis of 4-[(4-nitrophenoxy)carbonyl]phenyl 2,4-dimethoxybenzoate*

To a pre-dried flask flushed with argon and kept in an ice bath in order to maintain the temperature at 0°C, 4-(2,4-dimethoxybenzoyloxy)benzoic acid (2.00 g, 6.62×10-3 mol), 4-nitrophenol (1.01 g, 7.28×10-3 mol) and 4-dimethylaminopyridine (0.105 g, 8.60×10-4 mol) were added. The solids were solubilised with dichloromethane (50 mL) and stirred for 10 min before *N,N'*-dicyclohexylcarbodiimide (1.77 g, 8.60×10-3 mol) was added to the flask. The temperature of the reaction mixture was increased to room temperature and the reaction was allowed to proceed overnight. The white precipitate which formed was removed by vacuum filtration and the filtrate collected. The solvent was removed under vacuum and the crude product was purified using a silica gel column with 95 % dichloromethane and 5 % ethyl acetate as the eluent (RF value quoted in product data). The eluent fractions of interest were evaporated under vacuum to leave a white solid which was recrystallised from hot acetonitrile (50 mL).

Yield: 0.613 g, 21.9 %. RF: 0.58

$T_{Crl}$ 139 °C $T_{N_FN}$ (131 °C) $T_{NI}$ 188 °C

IR ($v_{max}$/cm$^{-1}$): 1742 (C=O ester)

$^1$H NMR (400 MHz, CDCl$_3$): 8.34 (2 H, d, J 9.1 Hz, Ar-H), 8.26 (2 H, d, J 8.8 Hz, Ar-H), 8.10 (1 H, d, J 8.7 Hz, Ar-H), 7.44 (2 H, d, J 9.1 Hz, Ar-H), 7.40 (2 H, d, J 8.8 Hz, Ar-H), 6.59 (1 H, dd, J 8.7 Hz, 2.4 Hz, Ar-H), 6.55 (1 H, d, J 2.4 Hz, Ar-H), 3.95 (3 H, s, Ar-O-CH$_3$), 3.91 (3 H, s, Ar-O-CH$_3$)

$^{13}$C NMR (101 MHz, CDCl$_3$): 165.39, 163.63, 162.79, 162.54, 155.99, 155.70, 145.44, 134.65, 131.89, 125.57, 125.30, 122.66, 122.52, 110.39, 104.99, 99.03, 56.07, 55.65

Data consistent with reported values.[1,2]

**Benzyl 2-fluoro-4-hydroxybenzoate**

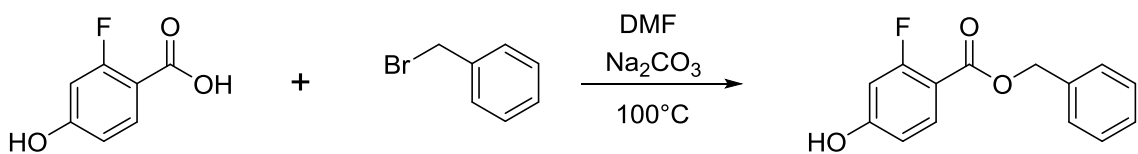

*Scheme 4 - Synthesis of 2-fluoro-4-hydroxybenzoate*

2-Fluoro-4-hydroxybenzoic acid (14.98 g, 96 mmol) and $Na_2CO_3$ (10.18 g, 96 mmol) were first azeotroped in dry toluene three times. Following this, dry DMF (100 ml) was added, and the mixture was heated to 100 °C. Benzyl bromide (16.42 g, 11.40 mL, 96 mmol) was added and the mixture was left stirring for 3 h. The mixture was added to 100 ml distilled water and acidified to approximately pH4 using dilute HCl. It was extracted using diethyl ether (5 x 25 mL). The combined organic extracts were washed using water and brine before finally being dried over magnesium sulfate. The crude product was purified by flash chromatography using petroleum ether (80%) and ethyl acetate (20%) as eluent (RF value quoted in product data). The residue was recrystalised from toluene (30 mL) to produce a white solid.

Yield: 5.83 g, 23 %. RF: 0.25. MP: 146 °C

IR ($v_{max}$/cm$^{-1}$): 3193 (alcoholic OH stretch), 2084 (aromatic overtones), 1665 (C=O stretching, ester)

$^1$H NMR (400 MHz, $C_2D_6OS$): 10.84 (1 H, s, OH), 7.79 (1 H, t, J 8.8 Hz, Ar-H), 7.48 – 7.29 (5 H, m, Ar-H), 6.75 – 6.60 (2 H, m, Ar-H), 5.30 (2 H, s, (C=O)-O-C$\underline{H}_2$-)

$^{13}$C NMR (101 MHz, $C_2D_6OS$): 164.22, 163.73, 163.61, 163.15 (d), 161.66, 136.27, 133.44 (d), 128.49, 128.00, 127.80, 112.06 (d), 108.52, 108.42, 103.72, 103.47, 65.80

**4-[(Benzyloxy)carbonyl]-3-fluorophenyl 2,4-dimethoxybenzoate**

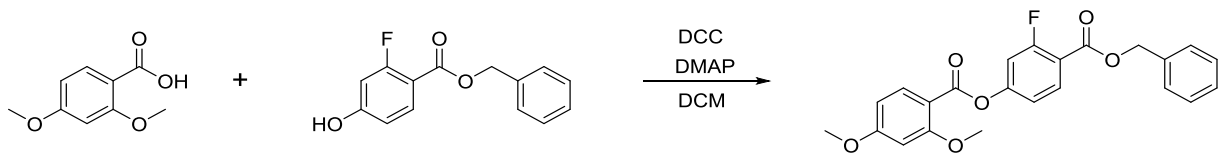

*Scheme 5 - Synthesis of 4-[(benzyloxy)carbonyl]-3-fluorophenyl 2,4-dimethoxybenzoate*

2,4-Dimethoxybenzoic acid (11.38 g, 62 mmol) and *N,N'*-dicyclohexylcarbodiimide (15.63 g, 75 mmol) were dissolved in DCM (250 mL) at 0 °C and left stirring for 10 min. Benzyl 2-fluoro-4-hydroxybenzoate (10.60 g, 43.1 mmol) and 4-dimethylaminopyridine (0.53 g, 4.31 mmol) were added together, and the mixture was left stirring at room temperature overnight. The mixture was filtered to remove the dicyclohexylurea. The crude product was purified by flash chromatography using petroleum ether (70%) and ethyl acetate (30%) as eluent to produce a white solid.

Yield: 11.14 g, 63% RF: 0.40 MP: 61 °C

IR ($v_{max}$/cm$^{-1}$): 2926 (C-H stretch), 1743 (C=O stretching, ester), 1734 (C=O stretching, ester)

$^1$H NMR (400 MHz, $C_2D_6OS$): 8.05 – 7.94 (2 H, m, Ar-H), 7.52 – 7.45 (2 H, m, Ar-H), 7.45 – 7.32 (4 H, m, Ar-H), 7.24 (1 H, d, J 8.4 Hz, Ar-H), 6.74 – 6.64 (2 H, m, Ar-H), 5.37 (2 H, s, (C=O)-O-C$\underline{H}_2$-), 3.89 (3 H, s, Ar-O-C$\underline{H}_3$), 3.85 (3 H, s, Ar-O-C$\underline{H}_3$)

$^{13}$C NMR (101 MHz, $C_2D_6OS$): 165.16, 162.77 (d), 161.96, 161.90, 160.19, 155.50, 155.38, 135.85, 134.20, 132.70 (d), 128.53, 128.15, 127.93, 118.78 (d), 115.43, 115.33, 111.72, 111.47, 109.31, 105.80, 99.38, 66.11, 56.03, 55.77

**4-[(2,4-Dimethoxybenzoyl)oxy]-2-fluorobenzoic acid**

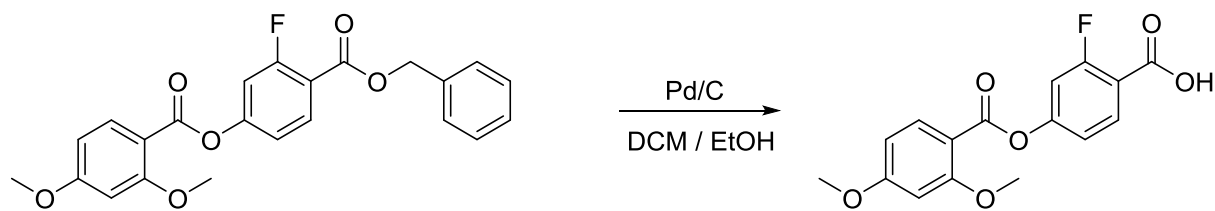

*Scheme 6 - Synthesis of 4-[(2,4-dimethoxybenzoyl)oxy]-2-fluorobenzoic acid*

4-[(Benzyloxy)carbonyl]-3-fluorophenyl 2,4-dimethoxybenzoate (11.00 g, 26.83 mmol) was dissolved in a 50:50 mixture of DCM and ethanol (300 mL). This solution was first evacuated under vacuum and purged with argon. Pd/C 5% (2.10 g, 6.72 mmol) was added, the argon atmosphere was replaced with hydrogen gas and the mixture was left stirring for 3 h. Hydrogen was pumped out of the system and the flask was purged thoroughly with argon. The mixture was filtered through Celite, and the solvents were removed under vacuum to produce a white solid which was used without further purification.

Yield: 3.61 g, 42 %. MP: 183 °C

IR ($v_{max}$/cm$^{-1}$): 3500 – 2500 (OH stretch, carboxylic acid), 2941 (C-H stretch), 2050 (aromatic overtones), 1688 (C=O stretching, carboxylic acid)

$^1$H NMR (400 MHz, C$_2$D$_6$OS): 13.29 (1 H, s, OH), 8.02 – 7.89 (2 H, m, Ar-H), 7.32 (1 H, d, J 11.3 Hz, Ar-H), 7.19 (1 H, d, J 6.5 Hz, Ar-H), 6.76 – 6.63 (2 H, m, Ar-H), 3.87 (6 H, s, Ar-O-<u>CH$_3$</u>)

$^{13}$C NMR (101 MHz, C$_2$D$_6$OS): 165.12, 164.50 (d), 162.77, 162.04, 161.92, 160.19, 154.88 (d), 134.17, 132.76 (d), 118.45 (d), 116.73 (d), 111.49, 111.23, 109.45, 105.79, 99.02, 56.04, 55.77

**Compound 2: 3-Fluoro-4-[(4-nitrophenoxy)carbonyl]phenyl 2,4-dimethoxybenzoate**

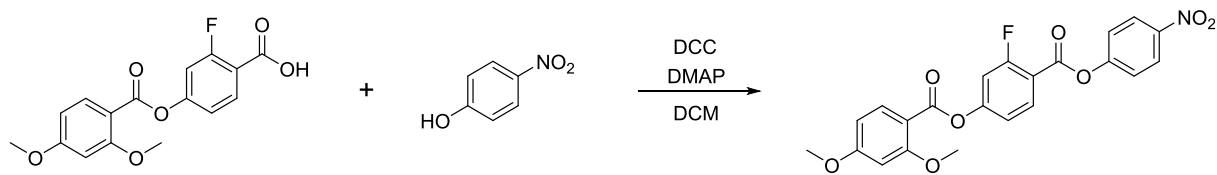

*Scheme 7 - Synthesis of 3-fluoro-4-[(4-nitrophenoxy)carbonyl]phenyl 2,4-dimethoxybenzoate*

4-[(2,4-Dimethoxybenzoyl)oxy]-2-fluorobenzoic acid (0.300 g, 0.94 mmol) and N,N'-dicyclohexylcarbodiimide (0.200 g, 1.12 mmol) were dissolved in DCM (50 mL) at 0 °C and left stirring for 10 min. 4-Nitrophenol (0.120 g, 0.85 mmol) and 4-dimethylaminopyridine (10 mg, 0.085 mmol) were added together, and the mixture was left stirring at room temperature overnight. The mixture was filtered to remove the dicyclohexylurea. The crude product was purified by flash chromatography using dichloromethane as eluent (RF value quoted in product data) to produce a white solid.

Yield: 0.100 g, 27 %. RF: 0.50

$T_{CrN}$ 161 °C $T_{N_FN}$ (143 °C) $T_{NI}$ 165 °C

M/Z: [M+Na]$^+$ Calculated mass for C$_{22}$H$_{16}$FNO$_8$Na: 464.0758. Found: 464.0741. Difference: -3.7 ppm

IR ($v_{max}$/cm$^{-1}$): 2941 (C-H stretch), 2115 (aromatic overtones), 1743 (C=O stretching, ester)

$^1$H NMR (400 MHz, C$_2$D$_6$OS): 8.37 (2 H, d, J 8.9 Hz, Ar-H), 8.22 (1 H, t, J 8.5 Hz, Ar-H), 8.00 (1 H, d, J 8.7 Hz, Ar-H), 7.65 (2 H, d, J 8.7 Hz, Ar-H), 7.50 (1 H, d, J 7.1 Hz, Ar-H), 7.34 (1 H, d, J 8.3 Hz, Ar-H), 6.76 – 6.64 (2 H, m, Ar-H), 3.89 (6 H, s, Ar-O-<u>CH$_3$</u>)

$^{13}$C NMR (101 MHz, C$_2$D$_6$OS): δ 165.70, 163.83, 162.51, 162.27, 161.23, 161.13 (d), 156.86, 156.74, 155.60, 145.78, 134.73, 133.90, 125.83, 123.88, 119.45 (d), 114.61 (d), 112.40, 112.15, 109.69, 106.30, 99.48, 56.52, 56.25

**Compound 3: 3-Fluoro-4-[(3-fluoro-4-nitrophenoxy)carbonyl]phenyl 2,4-dimethoxybenzoate**

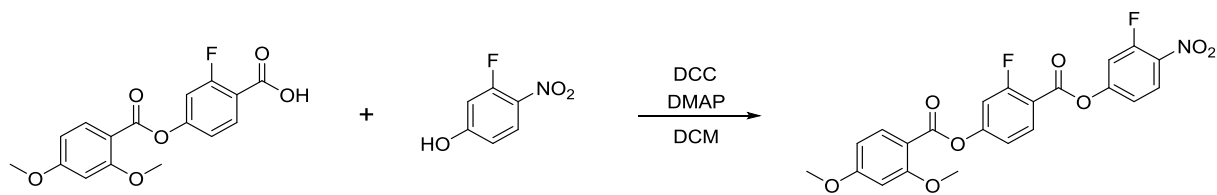

*Scheme 8 - Synthesis of 3-fluoro-4-[(3-fluoro-4-nitrophenoxy)carbonyl]phenyl 2,4-dimethoxybenzoate*

4-[(2,4-Dimethoxybenzoyl)oxy]-2-fluorobenzoic acid (0.600 g, 1.87 mmol) and N,N'-dicyclohexylcarbodiimide (0.460 g, 2.21 mmol) were dissolved in DCM (100 mL) at 0 °C and left stirring for 10 min. 3-Fluoro-4-nitrophenol (0.270 g, 1.72 mmol) and 4-dimethylaminopyridine (21 mg, 0.172 mmol) were added together, and the mixture was left stirring at room temperature overnight. The mixture was filtered to remove the dicyclohexylurea. The crude product was purified by flash chromatography using dichloromethane as eluent (RF value quoted in product data) to produce a white solid.

Yield: 0.190 g, 24 %. RF: 0.50

$T_{CrI}$ 151 °C $T_{N_FI}$ (139 °C)

M/Z: [M+Na]$^+$ Calculated mass for C$_{22}$H$_{15}$F$_2$NO$_8$Na: 482.0663. Found: 482.0669. Difference: 1.2 ppm

IR ($v_{max}$/cm$^{-1}$): 2978 (CH stretch), 2087 (aromatic overtones), 1742 (C=O stretching, ester)

$^1$H NMR (400 MHz, C$_2$D$_6$OS): 8.33 (1 H, t, J 8.8 Hz, Ar-H), 8.21 (1 H, t, J 8.5 Hz, Ar-H), 8.01 (1 H, d, J 8.8 Hz, Ar-H), 7.79 (1 H, d, J 11.8 Hz, Ar-H), 7.50 (2 H, dd, J 10.3 Hz, 7.2 Hz, Ar-H), 7.35 (1 H, d, J 8.6 Hz, Ar-H), 6.78 – 6.65 (2 H, m, Ar-H), 3.89 (6 H, s, Ar-O-<u>CH$_3$</u>)

¹³C NMR (101 MHz, C₂D₆OS): 165.25, 163.42, 162.05, 161.78, 160.82 (d), 160.26, 156.52, 156.41, 155.33, 155.14, 153.92, 134.27, 133.49, 127.59 (d), 119.28 (d), 119.03 (d), 113.94, 113.71, 113.01, 112.77, 111.98, 111.73, 109.19, 105.85, 98.65, 56.06, 55.79

**2-(3,5-Difluorphenyl)-5-propyl-1,3-dioxane**

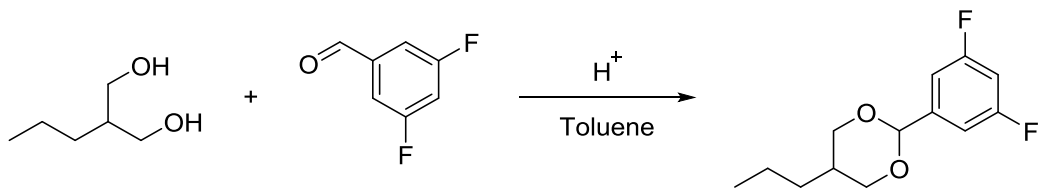

*Scheme 9 - Synthesis of 2-(3,5-difluorphenyl)-5-propyl-1,3-dioxane*

A clean dry round bottom flask was equipped with a stir bar, a dean stark device and a condenser. 2-Propyl-1,3-propanediol (1.47 g, 12.46 mmol) and 3,5-difluorobenzaldehyde (1.76 g, 12.46 mmol) were added and dissolved in toluene (140 mL). With stirring, *p*-toluenesulfonic acid (0.100 g, 0.58 mmol) was added. The mixture was kept under argon and refluxed stirring overnight. The mixture was cooled to room temperature, washed with brine followed by saturated sodium bicarbonate solution, dried over magnesium sulfate and concentrated under vacuum. The crude product was purified by flash column chromatography, using a 98:2 % mix of petroleum ether: ethyl acetate as eluent (RF value quoted in the product data) to produce a white solid.

Yield: 1.39 g, 46 %. RF: 0.48. MP: 28 °C

IR ($v_{max}$/cm⁻¹): 2931 (C-H stretch)

¹H NMR (400 MHz, CDCl₃): 7.07 – 6.98 (2 H, m, Ar-H), 6.77 (1 H, tt, J 8.9 Hz, 2.4 Hz, Ar-H), 5.36 (1 H, s, Ar-CH-), 4.23 (2 H, dd, J 11.5 Hz, 4.4 Hz, O-CH₂-CH-), 3.52 (2 H, t, J 11.2 Hz, -O-CH₂-CH-), 2.20 – 2.06 (1 H, m, -(CH₂)₂-CH-CH₂-), 1.39 – 1.29 (2 H, m, -CH-CH₂-CH₂-), 1.13 – 1.04 (2 H, m, -CH₂-CH₂-CH₃), 0.93 (3 H, t, J 7.3 Hz, -CH₂-CH₂-CH₃)

¹³C NMR (101 MHz, CDCl₃): 164.31, 163.79, 161.85, 161.72, 142.32 (t), 109.52, 109.45, 109.33, 109.26, 104.38, 104.13, 103.52, 99.79 (t), 72.69, 34.04, 30.41, 19.67, 14.33

Data consistent with reported values. ³,⁴

**2,3',4',5'-Tetrafluoro[1,1'-biphenyl]-4-ol**

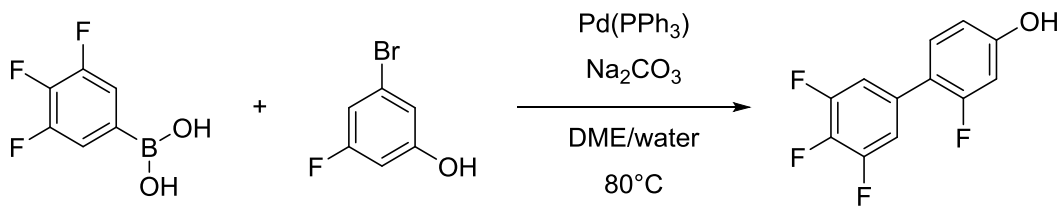

*Scheme 10 - Synthesis of 2,3',4',5'-tetrafluoro[1,1'-biphenyl]-4-ol*

A round bottom flask was equipped with a stir bar and a condenser, it was charged with 4-bromo-3-fluorophenol (1.31 g, 6.87 mmol), 3,4,5-trifluoroboronic acid (1.45 g, 8.24 mmol) and sodium carbonate (1.82 g, 17.18 mmol) in a dimethoxyethane: water mix (11.1 mLl: 5.5mL). The mixture was sparged with argon for 30 min before Pd(PPh₃) (0.240 g, 0.208 mmol) was added. The mixture was kept under argon and stirred at 80°C overnight. The mixture was cooled to room temperature and added to toluene. It was acidified using concentrated HCl, washed with water, filtered over Celite, dried over magnesium sulfate, and concentrated under vacuum. The crude product was purified by flash column chromatography, using an 80:20 % mix of petroleum ether: ethyl acetate as eluent (RF value quoted in the product data). The recovered residue was recrystalised from hexane (10 mL) to produce a white solid.

Yield: 1.03 g, 62 %. RF: 0.30. MP: 96 °C

IR ($v_{max}$/cm⁻¹): 3315 (alcoholic OH stretch)

¹H NMR (400 MHz, CDCl₃): 7.23 (1 H, t, J 8.1 Hz, Ar-H), 7.17 – 7.07 (2 H, m, Ar-H), 6.76 – 6.61 (2 H, m, Ar-H), 5.48 (1 H, s, OH)

¹³C NMR (101 MHz, CDCl₃): 161.33, 158.86, 157.24, 157.13, 152.34 (dd), 149.85 (dd), 140.26, 137.76, 130.83 (d), 118.68 (d), 112.81 (m), 111.98 (d), 104.09, 103.84

Data consistent with reported values. ³,⁴

**2,6-Difluoro-4-(5-propyl-1,3-dioxan-2-yl) benzoic acid**

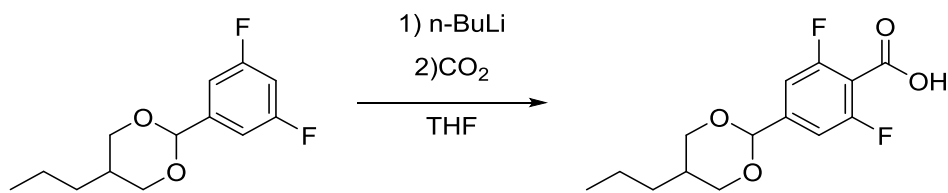

*Scheme 11 - Synthesis of 2,6-difluoro-4-(5-propyl-1,3-dioxan-2-yl) benzoic acid*

A clean dry two necked round bottom flask equipped with a stir bar and rubber septum was charged with 2-(3,5-difluorphenyl)-5-propyl-1,3-dioxane (0.930 g, 3.82 mmol) under argon via syringe in toluene. The toluene was removed under vacuum, and this was repeated two more times. Freshly distilled THF (30 mL) was added via syringe, and this mixture was kept cold by using a dry ice / acetone bath. After allowing the mixture to cool *n*-BuLi (0.310 g, 2.98 mL, 4.80 mmol) was cautiously added dropwise via syringe. The mixture was kept cold stirring for 1 h, after this dry ice was added directly as a solid in excess. This was left stirring until the mixture de-gassed, noted by the bubbles slowing to a constant rate in the oil bubbler connected to the argon gas and reaction flask. The reaction was quenched carefully with distilled water. The mixture was acidified with concentrated HCl and THF was removed under vacuum. The resulting aqueous layer was extracted several times with diethyl ether, these combined organic extracts were washed with water to remove the valeric acid (formed as a side product from unreacted *n*-BuLi upon adding dry ice), presence indicated by an unpleasant smell. It was dried over magnesium sulfate to produce a white solid which was used without further purification.

Yield: 0.930 g, 85 %. MP: 125 °C

IR ($v_{max}$/cm$^{-1}$): 3500 – 2500 (OH stretch, carboxylic acid), 2928 (C-H stretch), 2000 (aromatic overtones), 1701 (C=O stretching, carboxylic acid)

$^1$H NMR (400 MHz, CDCl$_3$): 7.13 (2 H, d, J 9.5 Hz, Ar-H), 5.37 (1 H, s, Ar-CH-), 4.24 (2 H, dd, J 11.6 Hz, 4.5 Hz, O-CH$_2$-CH-), 3.52 (2 H, t, J 11.5 Hz, -O-CH$_2$-CH-), 1 H, m, -(CH$_2$)$_2$-CH-CH$_2$-), 1.38 – 1.29 (2 H, m, -CH-CH$_2$-CH$_2$-), 1.14 – 1.05 (2 H, m, -CH$_2$-CH$_2$-CH$_3$), 0.93 (3 H, t, J 7.3 Hz, -CH$_2$-CH$_2$-CH$_3$)

$^{13}$C NMR (101 MHz, CDCl$_3$): 165.14, 162.50 (d), 159.94 (d), 145.39 (t), 110.37 (d), 110.14 (d), 98.84, 72.56, 33.87, 30.67, 19.52, 14.19

Data consistent with reported values. [3,4]

**Compound 4: 2,3',4',5'-Tetrafluoro[1,1'-biphenyl]-4-yl-2,6-difluoro-4-(5-propyl-1,3-dioxan-2-yl) benzoate (DIO)**

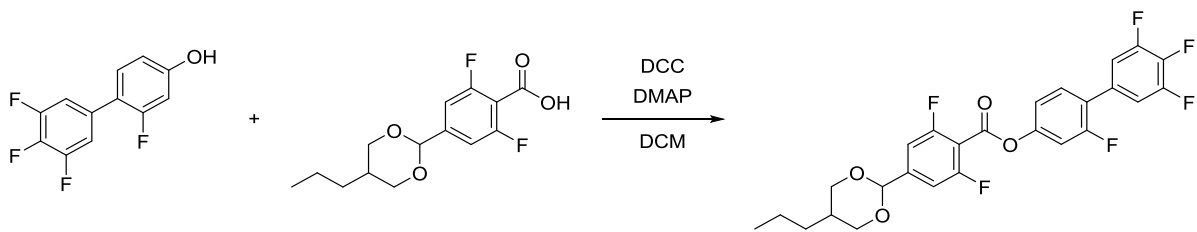

*Scheme 12 - Synthesis of 2,3',4',5'-tetrafluoro[1,1'-biphenyl]-4-yl-2,6-difluoro-4-(5-propyl-1,3-dioxan-2-yl) benzoate*

A clean dry round bottom flask equipped with a stir bar was charged with 2,3',4',5'-tetrafluoro[1,1'-biphenyl]-4-ol (0.300 g, 1.26 mmol), 2,6-difluoro-4-(5-propyl-1,3-dioxan-2-yl) benzoic acid (0.520 g, 1.83 mmol) and *N,N'*-dicyclohexylcarbodiimide (0.450 g, 2.22 mmol) in DCM (50 mL) at 0°C. This mixture was left stirring for 30 min before 4-dimethylaminopyridine (15 mg, 0.123 mmol) was added. The mixture was heated to room temperature and left stirring under argon overnight. The mixture was filtered to remove the dicyclohexylurea formed during the reaction. The DCM was removed under vacuum and the residue was purified by flash column chromatography using a 95:5 % mix of petroleum ether: ethyl acetate as eluent (RF value quoted in the product data). The residue was recrystalised from hexane (10 mL) to produce a white solid.

Yield: 0.430 g, 67 %. RF: 0.23

T$_{CrN}$ 96 °C T$_{N_FN_X}$ (68 °C) T$_{N_XN}$ (84 °C) T$_{NI}$ 174 °C

M/Z: [M+Na]$^+$ Calculated mass for C$_{26}$H$_{20}$F$_6$O$_4$: 533.1163. Found: 533.1165. Difference: 0.4 ppm

IR ($v_{max}$/cm$^{-1}$): 2926 (C-H stretch), 2000 (aromatic overtones), 1738 (C=O stretching, ester)

$^1$H NMR (400 MHz, CDCl$_3$): 7.43 (1 H, t, J 8.6 Hz, Ar-H), 7.24 – 7.13 (6 H, m, Ar-H), 5.40 (1 H, s, Ar-CH-), 4.26 (2 H, dd, J 11.8 Hz, 4.6 Hz, O-CH$_2$-CH-), 3.54 (2 H, t, J 11.4 Hz, -O-CH$_2$-CH-), 2.21 – 2.07 (1 H, m, -(CH$_2$)$_2$-CH-CH$_2$-), 1.42 – 1.27 (2 H, m, -CH-CH$_2$-CH$_2$-), 1.16 – 1.05 (2 H, m, -CH$_2$-CH$_2$-CH$_3$), 0.94 (3 H, t, J 7.3 Hz, -CH$_2$-CH$_2$-CH$_3$)

$^{13}$C NMR (101 MHz, CDCl$_3$): 162.21 (d), 160.61, 159.64 (d), 159.29 (t), 158.12, 152.43 (dd), 151.02, 150.92, 149.95 (dd), 145.64 (t), 140.80 (t), 138.28 (t), 130.66 (d), 124.31 (d), 118.14 (d), 113.21 (m), 110.78, 110.52, 110.43 (d), 110.19 (d), 109.43, 98.80 (t), 71.29, 33.90, 28.65, 18.06, 13.81

# Experimental methods

## Calorimetric Measurements

Calorimetric studies were performed with a TA DSC Q200 calorimeter, 1 - 3 mg samples were sealed in aluminum pans and kept in nitrogen atmosphere during measurement, and both heating and cooling scans were performed with a rate of 5–10 K/min.

## Microscopic Studies

Optical studies were performed by using the Zeiss Axio Imager A2m polarized light microscope, equipped with Linkam heating stage. Samples were prepared in commercial cells (AWAT) of various thickness (1.5 – 20 μm) having ITO electrodes and surfactant layer for either planar or homeotropic alignment, HG and HT cells, respectively.

## X-Ray Diffraction

The wide angle X-ray diffraction patterns were obtained with the Bruker D8 GADDS system (CuKα radiation, Goebel mirror monochromator, 0.5 mm point beam collimator, Vantec2000 area detector), equipped with modified Linkam heating stage. Samples were prepared as droplets on a heated surface.

For single crystal x-ray measurements a specimen of compound 4 was mounted on a kapton loop with a drop of ParatoneN oil. Intensity data from single crystal X-ray diffraction were measured on Rigaku Oxford Diffraction Supernova 4 circle diffractometer equipped with copper (CuKα) microsource and Atlas CCD detector at 120K. The temperature of the sample was controlled with a precision of ± 0.1 K using Oxford Cryosystems cooling device. The data were collected, integrated and scaled with CrysAlis171[1] software. The structures were solved by direct methods using SXELXT[2] and refined by full-matrix least squares procedure with SHELXL[3] within OLEX2[4] graphical interface. Structure was deposited with CCDC (deposition number 2098521) and can be retrieved upon request. The most important crystallographic parameters are collected in Table S1.

## Birefringence Measurements

Birefringence was measured with a setup based on a photoelastic modulator (PEM-90, Hinds) working at a modulation frequency f = 50 kHz; as a light source, a halogen lamp (Hamamatsu LC8) was used, equipped with a narrow bandpass filter (532 nm). The signal from a photodiode (FLC Electronics PIN-20) was de-convoluted with a lock-in amplifier (EG&G 7265) into 1f and 2f components to yield a retardation induced by the sample. Knowing the sample thickness, the retardation was recalculated into optical birefringence. Uniformly aligned samples in HG cells, 1.5- and 3-μm-thick, were measured.

## Spontaneous Electric Polarization Measurements

Values of the spontaneous electric polarization were obtained from the current peaks recorded during Ps switching upon applying triangular voltage at a frequency of 20 – 200 Hz. The 20-μm-thick cells with ITO electrodes and HT anchoring were used, switching current was determined by recording the voltage drop at the resistivity of 1kΩ in serial connection with the sample. The current peak was integrated over time to calculate the surface electric charge and evaluate polarization value.

## Dielectric spectroscopy

The complex dielectric permittivity, $\varepsilon^*$, was measured in 1 Hz – 10 MHz frequency ($f$) range using Solatron 1260 impedance analyzer. Material was placed in glass cells with ITO or Au electrodes and thickness ranging from 1.5 to 75 microns. The relaxation frequency, $f_r$, and dielectric strength of the mode, $\Delta\varepsilon$, were evaluated by fitting complex dielectric susceptibility to the Cole-Cole formula:

$$\varepsilon^* - \varepsilon_\infty = \sum \frac{\Delta\varepsilon}{1+(if/f_r)^{1-\alpha}} + i\frac{\sigma}{2\pi\varepsilon_0 f}$$

where $\varepsilon_\infty$, $\alpha$ and $\sigma$ are high frequency dielectric constant, distribution parameter of the mode and low frequency conductivity, respectively. Example of fitting is presented in Fig. S8.

## Splay elastic constant measurements

All the materials exhibit a positive dielectric anisotropy over the whole temperature range of the N phases; thus the elastic constant measurements were performed in cells with planar alignment, in which an electric field was applied across the cell. The dielectric permittivity was measured with the Wayne Kerr Precision Component Analyzer 6425, as a function of the applied voltage amplitude (*V*) ranging from 0.01 to 5.0 V, at the frequency 12 kHz. The splay elastic constant $K_{11}$ was determined from the threshold voltage $V_{th}$ at which the director reorientation starts and thus the effective permittivity ($\varepsilon$) starts to grow, as $K_{11} = \Delta\varepsilon\varepsilon_0(V^2{}_{th}/\pi^2)$.

## Piezoresponse Force Microscopy

PFM experiments were performed in optimized vertical domains mode, in this mode deflection of conducting AFM cantilever is detected when the piezoelectric material domains mechanically deform upon applied voltage. Because the magnitude of displacement is very small, a lock-in technique is used, where a modulated voltage reference signal is applied to the tip which causes deformation of the sample surface. The AFM tip is maintained in contact mode. The lock-in amplifier measures the cantilever deflection signal component which during deformation oscillates in phase with the drive amplitude (reference) signal and out of phase, perpendicular to the drive amplitude signal, the signals are re-calculated into the amplitude and phase angle of the cantilever deflection. In ideal situation at the domain wall the amplitude drops to zero and phase changes by 180 degree.

The sample was placed on the ITO electrode surface, which was electrically connected to the AFM stage. PFM was measured on Bruker Dimension Icon using Bruker SCM-PIT cantilevers. Set point amplitude was 0.2V and drive amplitude was 8V. The presented pictures were measured at 0V tip bias.

**Modelling of Molecular Structure**

The geometric parameters and electronic properties of each compound were determined using quantum mechanical DFT calculations with Gaussian09 software.[5] Geometry optimisation was carried out at the B3LYP/6-31G(d) level of theory. In order to determine the lowest energy conformer for each compound, multiple structures were compared with the lateral substituents in equivalent positions on the aromatic rings, namely the 3 and 5 positions for the fluorine substituents, and the 2 and 6 positions for the methoxy group. For visualisation of space-filling models, QuteMol was used [6], and for visualisation of electrostatic potential surfaces, ball-and-stick models and dipole moments, GaussView 5 was used.[7]

## Results and Discussion

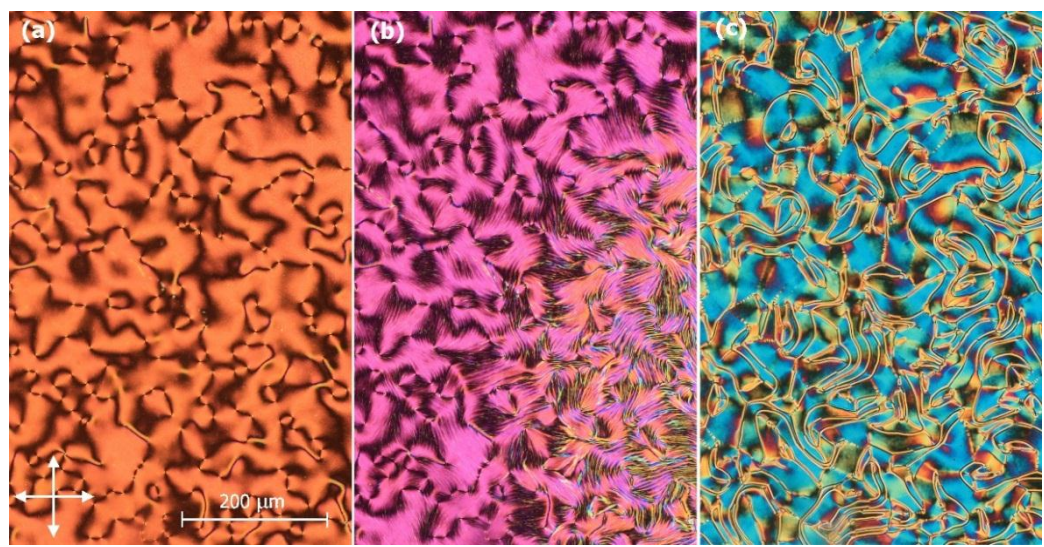

**Figure S1.** Optical textures of compound **2** (a) in N phase, (b) at N-$N_F$ phase transition and (c) in $N_F$ phase in a 3-μm-thick cell with homeotropic anchoring condition. In all panels the same area of the sample is shown.

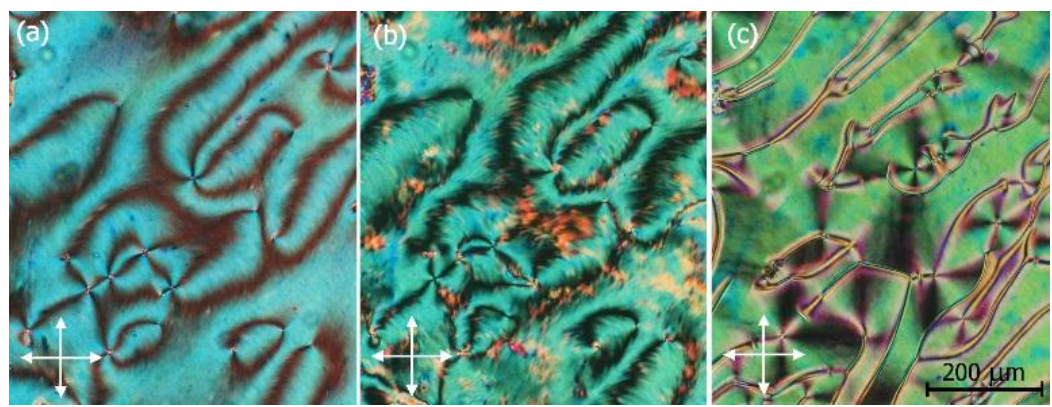

**Figure S2.** Optical textures of compound **4** in (a) N, (b) $N_x$ and (c) $N_F$ phases in a 3-μm-thick cell with homeotropic anchoring condition. In all panels the same area of the sample is shown.

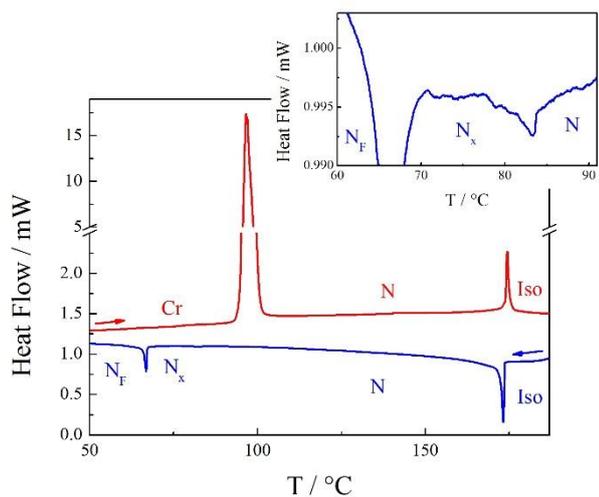

**Figure S3.** *DSC traces recorded for* compound **4** on heating (red line) and cooling (blue line).

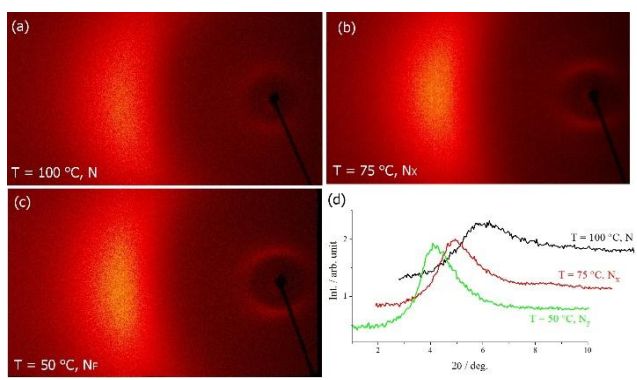

**Figure S4.** 2D X-ray diffraction patterns for compound **4** in (a) N, (b) Nx and (c) NF phases. (d) X-ray intensity vs. diffraction angle (2θ) obtained by azimuthal integration of low angle range of 2D x-ray patterns.

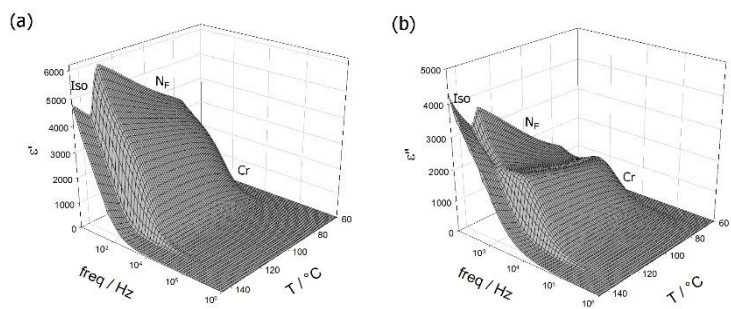

**Figure S5.** Dielectric dispersion for compound **3** measured in 20-μm-thick cell with homeotropic anchoring: (a) real and (b) imaginary part of dielectric susceptibility vs. temperature and frequency.

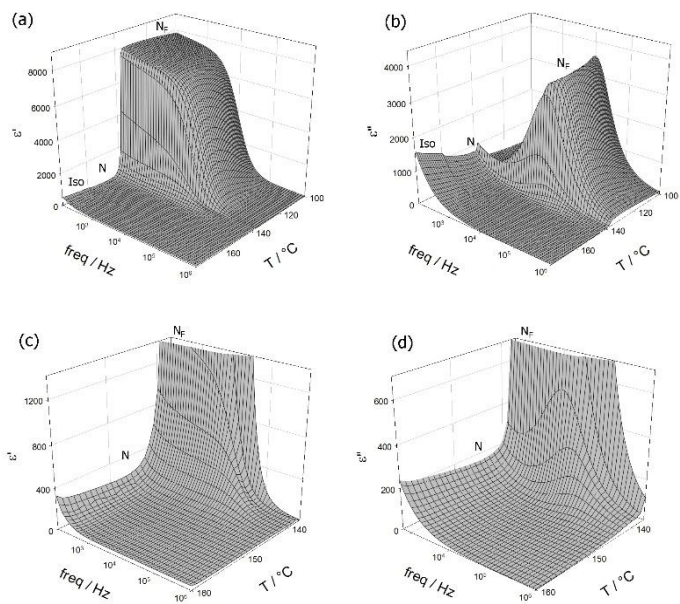

**Figure S6.** Dielectric dispersion for compound **2** measured in 5-μm-thick cell with gold electrodes: (a) real and (b) imaginary part of dielectric susceptibility vs. temperature and frequency, in (c) and (d) enlarged parts showing temperature region close to N-N$_F$ phase transition.

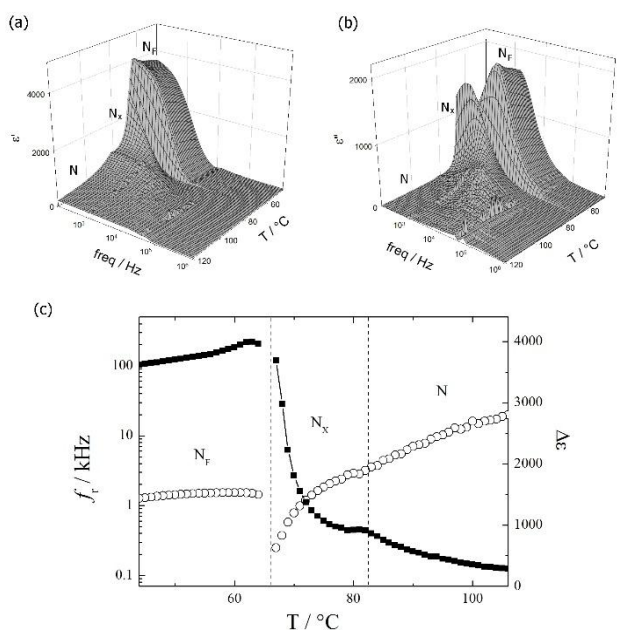

**Figure S7.** Dielectric dispersion for compound **2** measured in 20-μm-thick cell with homeotropic anchoring under bias electric field (0.25 V μm$^{-1}$): (a) real and (b) imaginary part of dielectric susceptibility vs. temperature and frequency, (c) relaxation frequency (open circles) and dielectric mode strength (solid squares) evaluated from above data by fitting to Cole-Cole formula.

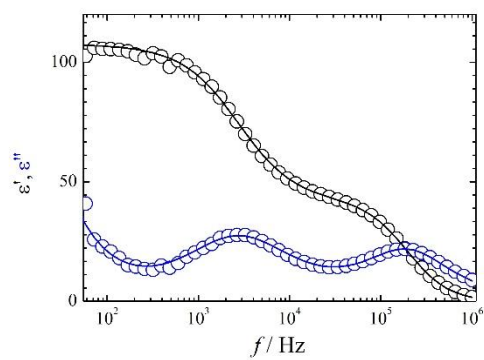

**Figure S8.** Example of the simultaneous fitting of real (black) and imaginary (blue) parts of dielectric susceptibility (measured for compound **4** in 20-μm-thick cell with homeotropic anchoring in $N_X$ phase at 75 °C) to the Cole-Cole formula with two relaxation modes, circles present measured data and lines are simulated curves. Fitting parameters of relaxation modes: $f_{r1}$=2749 Hz, $\Delta\varepsilon_1$=66.6, $\alpha_1$=0.15, $f_{r2}$=196273 Hz, $\Delta\varepsilon_2$=41.6, $\alpha_2$=0.02.

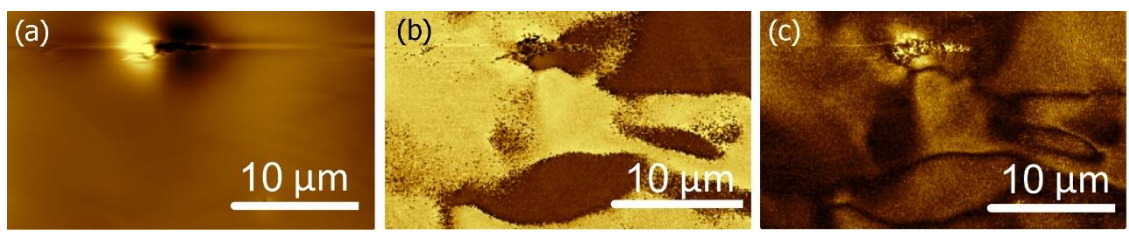

**Figure S9.** PFM images of compound **1** in NF phase: topology recorded as height sensor signal (a), polar domain structure recorded as a piezo-response signals (b) phase – yellow and brown colors show the opposite vertical orientation of polarization vector and (c) amplitude – color intensity shows the magnitude of vertical polarization component.

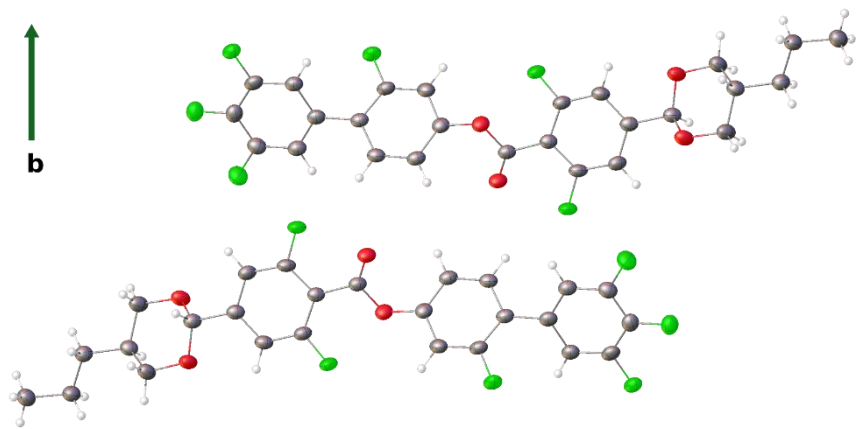

**Figure S9.** Molecular structure and arrangement of molecules in crystallographic unit cell determined crystal data and structure refinement from for compound **4**

**Table S1** Crystal data and structure refinement for **4**.

| | |
|---|---|
| Identification code | **4** |
| CCDC depostion # | 2098521 |
| Empirical formula | $C_{26}H_{20}F_6O_4$ |
| Formula weight | 510.42 |
| Temperature/K | 120.00(10) |
| Crystal system | monoclinic |
| Space group | $P2_1/c$ |
| a/Å | 16.3938(9) |
| b/Å | 12.8183(5) |
| c/Å | 10.7853(6) |
| α/° | 90 |
| β/° | 93.610(4) |
| γ/° | 90 |
| Volume/Å$^3$ | 2261.9(2) |
| Z | 4 |
| $\rho_{calc}$ g/cm$^3$ | 1.499 |
| μ/mm$^{-1}$ | 1.159 |
| F(000) | 1048.0 |
| Crystal size/mm$^3$ | 0.331 × 0.086 × 0.037 |
| Radiation | Cu Kα (λ = 1.54184) |
| 2Θ range for data collection/° | 5.402 to 145.642 |
| Index ranges | -18 ≤ h ≤ 20, -14 ≤ k ≤ 15, -13 ≤ l ≤ 11 |
| Reflections collected | 16033 |
| Independent reflections | 4418 [$R_{int}$ = 0.0417, $R_{sigma}$ = 0.0331] |
| Data/restraints/parameters | 4418/0/326 |
| Goodness-of-fit on F$^2$ | 1.035 |
| Final R indexes [I>=2σ (I)] | $R_1$ = 0.0449, $wR_2$ = 0.1145 |
| Final R indexes [all data] | $R_1$ = 0.0567, $wR_2$ = 0.1256 |
| Largest diff. peak/hole / e Å$^{-3}$ | 0.25/-0.29 |

**Full crystal data in file DIO_crystal_stucture.cif**

Authors thank Dr. Vladimira Novotna and Dr. Ladislav Fekete from Institute of Physics, Academy of Sciences in Czech Republic in Prague for their help in PFM measurements.